\pdfoutput=1
\documentclass[preprint,1p]{elsarticle}

\usepackage{amsfonts}
\usepackage{amssymb}
\usepackage{graphicx}
\usepackage{epstopdf}
\usepackage{verbatim}
\usepackage{amsmath}
\usepackage{wasysym}
\usepackage[thickspace,amssymb]{SIunits}

\begin{document}

\begin{frontmatter}

\title{Study of the time and space distribution of $\beta^+$ emitters from $80\ \mega\electronvolt/$u carbon ion beam irradiation on PMMA.}

\author{C.~Agodi$^{f}$, F.~Bellini$^{a,b}$, G.A.P.~Cirrone$^{f}$, F.~Collamati$^{a,b}$, G.~Cuttone$^{f}$, E.~De~Lucia$^{c}$, M.~De~Napoli$^{f}$, A.~Di~Domenico$^{a,b}$, R.~Faccini$^{a,b}$, F.~Ferroni$^{a,b}$, S.~Fiore$^{a}$, P.~Gauzzi$^{a,b}$, E.~Iarocci$^{c,d}$, M.~Marafini$^{e}$, I.~Mattei$^{g,c}$, A.~Paoloni$^{c}$, V.~Patera$^{c,d}$, L.~Piersanti$^{c,d}$, F.~Romano$^{e,f}$, A.~Sarti$^{c,d}$, A.~Sciubba$^{c,d}$, C.~Voena$^{a,b}$
}
\address{

$^a$ Dipartimento di Fisica, Sapienza Universit\`a di Roma, Roma, Italy \\
$^b$ INFN Sezione di Roma, Roma, Italy \\
$^c$ Laboratori Nazionali di Frascati dell'INFN, Frascati, Italy\\ 
$^d$ Dipartimento di Scienze di Base e Applicate per l'Ingegneria, Sapienza Universit\`a di Roma,  Roma, Italy\\
$^e$ Museo Storico della Fisica e Centro Studi e Ricerche ``E.~Fermi'', Roma, Italy\\
$^f$ Laboratori Nazionali del Sud dell'INFN, Catania, Italy
$^g$ Dipartimento di Fisica, Roma Tre Universit\`a di Roma, Roma, Italy \\
 
}
 
\begin{abstract}

Proton and carbon ion therapy is an emerging technique used for the treatment of solid cancers. The monitoring of the dose delivered during such treatments and the  on-line knowledge of the Bragg peak position is still a matter of research. 
A possible technique exploits the collinear $511\ \kilo\electronvolt$ photons produced by positrons annihilation from $\beta^+$ emitters created by the beam.
This paper reports rate measurements of the $511\ \kilo\electronvolt$ photons emitted after the interactions of a $80\ \mega\electronvolt / u$ fully stripped carbon ion beam  at the Laboratori Nazionali del Sud (LNS) of INFN, with a Poly-methyl methacrylate target. 
The time evolution of the $\beta^+$ rate was parametrized and the dominance of $^{11}C$ emitters over the other species ($^{13}N$,  $^{15}O$,  $^{14}O$) was observed, measuring the fraction of carbon ions activating $\beta^+$ emitters to be $(10.3\pm0.7)\cdot10^{-3}$.  The average depth in the PMMA of the positron annihilation from $\beta^+$ emitters was also measured, $D_{\beta^+}=5.3\pm1.1\ \milli\meter$,  to be compared to the expected Bragg peak depth $D_{Bragg}=11.0\pm 0.5\ \milli\meter$ obtained from simulations.

\end{abstract}


\end{frontmatter}

\section*{Introduction}

\begin{figure}[!ht]
\begin{center}
\includegraphics [width =0.7 \textwidth] {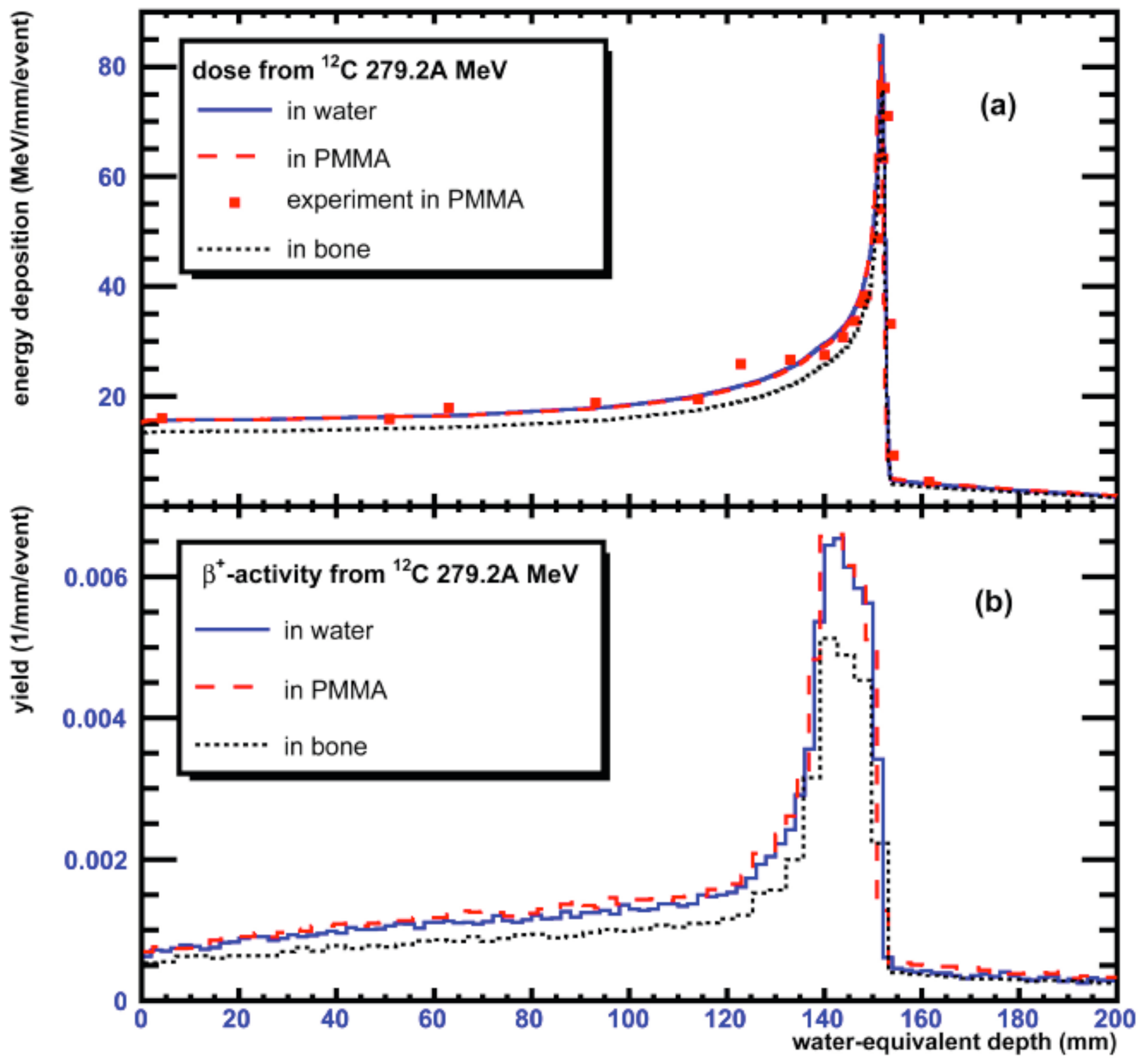}
\caption{\small{(a) Dose distributions as functions of water-equivalent depth estimated with the GEANT MC for $279.2\ \mega\electronvolt$ $^{12}C$ nuclei in water (solid-line histogram), PMMA (dashed histogram) and bone (dotted histogram). Experimental data for depth dose distributions in PMMA are shown by points~\protect \cite{Bragg}. (b) Distributions of positron-emitting nuclei produced in these materials as a function of water-equivalent depth~\protect \cite{Bragg}.}}
\label{fig:Beta}
\end{center}
\end{figure}

The use of proton and carbon ion beams has become more and more widespread as an effective therapy for the treatment of solid cancers (hadrontherapy). These beams have maximum energy density released at the Bragg peak (Fig.~\ref{fig:Beta} Top) at the end of their range, as opposed to X-rays or $\gamma$-rays, which are absorbed by the body with an exponential decrease in the delivered dose with increasing tissue depth after an initial build-up~\cite{Bragg}. 

Due to their very favorable profile of released dose in tissue, the charged hadron beams can be very effective in destroying the tumor and sparing the adjacent healthy tissue in comparison to the standard X-ray based treatment~\cite{Amaldi}.  On the other hand, the higher spatial selectivity of hadrontherapy asks for a dedicated approach to the delivered dose monitoring.

The uncertainty on the position of the dose release in hadrontherapy treatment can be due to different factors, i.e. quality and calibration of the Computed Tomography (CT) images,  possible morphologic changes occurring between CT and treatment,  patient mis-positioning and  organ motion during the treatment itself. All these effects give an overall uncertainty of the order of few millimeters, that can be larger than the dimension of the dose release spot at the Bragg peak. 

Several methods have been developed to determine the Bragg peak position by exploiting the secondary particle production induced by the charged hadron beam, and among these one of the most promising is the PET-like technique: the collinear $511\ \kilo\electronvolt$ photons produced by positrons annihilation from $\beta^+$ emitters created by the beam are measured. The relationship between the $\beta^+$ emitters densities and the dose release has been studied with Monte Carlo (MC) simulations, as shown in Fig.~\ref{fig:Beta}~\cite{Bragg}. The measurement of the rates of such emitters can also provide precise monitoring of the dose, which is in turn essential for a good quality control of the treatment. This technique has already been developed with measurements both after the irradiation~\cite{inBeam:th,inBeam:C} and  on-beam~\cite{inBeam:exp, rates2009}.

To this aim, measurements are needed to allow the MC tuning, which is critical for the appropriate development of the technique. In particular, a determination of the isotopic composition of the $\beta^+$ emitters and the corresponding rates has not been performed for the carbon ion treatments. Furthermore, papers in literature mostly report  time integrated measurements and do not investigate the time structure of the emission in presence of irradiation.

In this paper we present measurements of the properties of the $\beta^+$ emitters,  produced in a Poly-Methyl Methacrylate (PMMA) phantom during an irradiation with carbon ions, by observing the two $511\ \kilo\electronvolt$ photons produced in the positron annihilation ($\gamma-$PET). In section~\ref{Catania}, we describe the setup of the experiment performed at the INFN Laboratori Nazionali del Sud (LNS) in Catania in the interaction of a $80\ \mega\electronvolt/$u fully stripped carbon beam with a PMMA target (see~\cite{PaperoLYSO} for a related  experiment with the same setup).  Data analysis tools are detailed in Sec.~\ref{Eff}.

With the collected data we investigated three aspects of the $\beta^+$ emission: the isotopic composition of the emitters (Sec.~\ref{sec:isotopic}), their corresponding rates, via a study of the  time dependence of the emissions during irradiation (Sec.~\ref{tdep}), and the position of the $\beta^+$ emitters with respect to the Bragg peak (Sec.~\ref{Bragg}). 
With respect to the existing studies we have allowed for incident beam intensity variations.

\section{Experimental setup}
\label{Catania}

\begin{figure}[!ht]
\begin{center}
\includegraphics [width = 0.6 \textwidth] {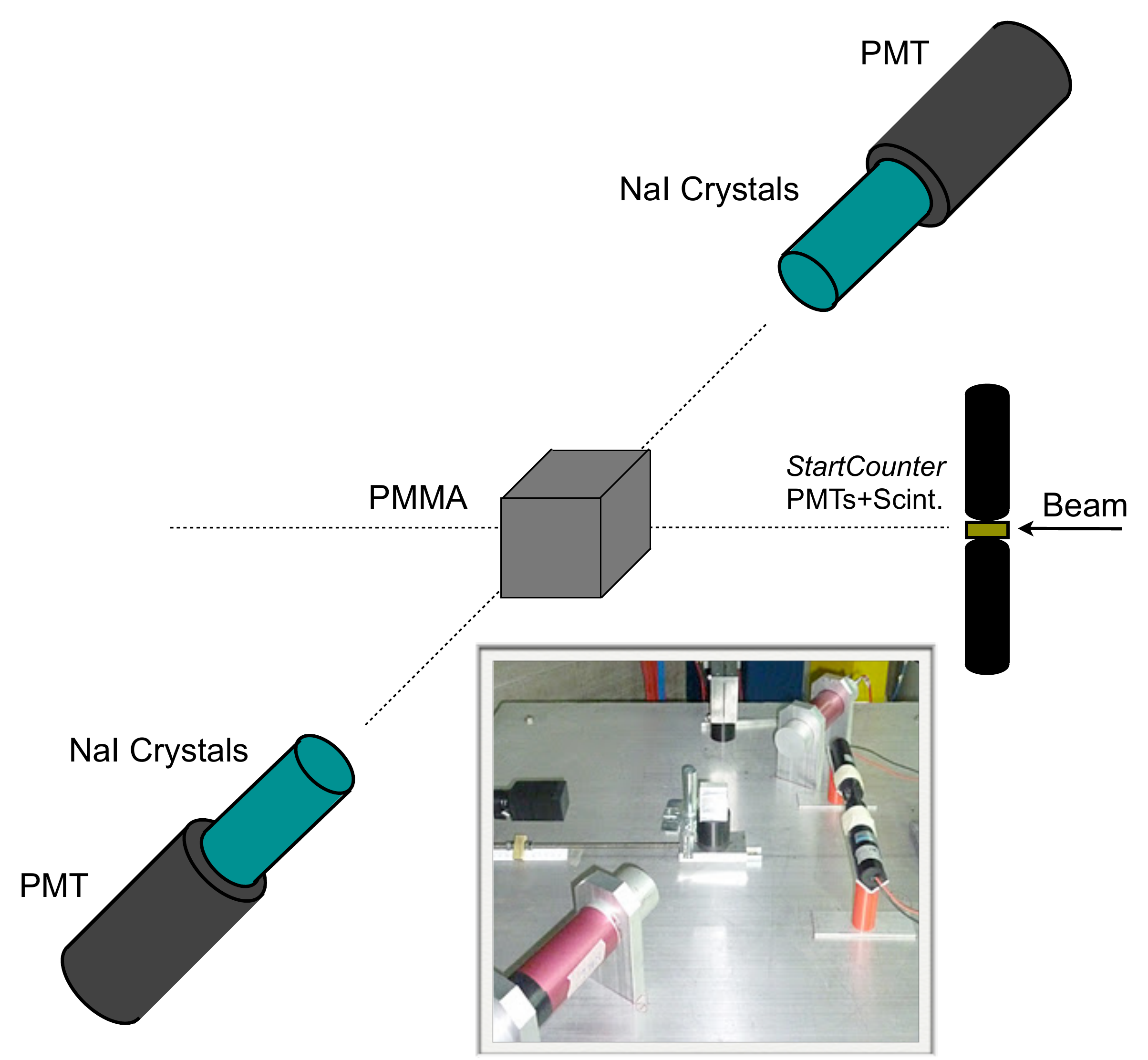}
\caption{\small{Schematic view of the experimental setup: the NaI detectors are placed $20\ \centi\meter$ away from the PMMA. The acquisition is triggered by the coincidence of the two crystals. On the right the picture of the actual experimental setup is shown. }}
\label{fig:Schema}
\end{center}
\end{figure}

The experimental setup is shown in Fig.~\ref{fig:Schema}. A $4\times 4 \times 4\ \centi\meter^3$ PMMA target is placed on an $80\ \mega\electronvolt/$u fully stripped $^{12}C$ ion beam. The beam rate, ranging from hundred of $\kilo\hertz$ to $\sim 2 \mega\hertz$, is monitored with a $1.1\ \milli\meter$ thick scintillator on the beam line, placed at  $17\ \centi\meter$ from the PMMA, and read-out by two photomultiplier tubes (PMTs) (Hamamatsu $10583$) put in coincidence (Start Counter).

A pair of cylindrical sodium iodide activated with thallium crystals NaI(Tl) ($r= 2.5\ \centi\meter$ and $h= 5 \ \centi\meter$) is placed at $45\degree$ ($225\degree$) with respect to the beam line, at $20\ \centi\meter$ from the PMMA. The scintillation light of the two crystals is detected by two Scionix $V14-$EI PMTs triggered in coincidence within a time window of $80\ \nano\second$. A $12$-bit QDC (Caen $V792$N) and a $19$-bit TDC (Caen $V1190$B) provide the measurements of energy and arrival time.
We use NaI crystals to detect the $\gamma-$PET signals because of their high light yield and energy resolution in the $O(1\ \mega\electronvolt )$ ranges.
In order to perform position-dependent measurements, the PMMA can be moved along the beam axis  with an accuracy of $0.5\ \milli\meter$.

\section{Data selection}
\label{Eff}

From the charge collected in the two NaI detectors we measure the energy of the two photons ($E_{\gamma}^i$, $i=1,2$) from the positron annihilation. To calibrate the NaI detectors we used $^{22}Na$ ($511 \ \kilo\electronvolt$), $^{137}Cs$ ($662 \ \kilo\electronvolt$) and $^{60}Co$ ($1.17 \ \mega\electronvolt$ and $1.33\ \mega\electronvolt$) sources. A good linearity is obtained up to $1.5\ \mega\electronvolt$. The relative energy resolution at $511\ \kilo\electronvolt$ is $\sigma_E/E = 2.7\%$; we define the signal window $0.48<E_i<0.53\ \mega\electronvolt$. 

\begin{figure}[!ht]
\begin{center}
\includegraphics [width =0.47 \textwidth] {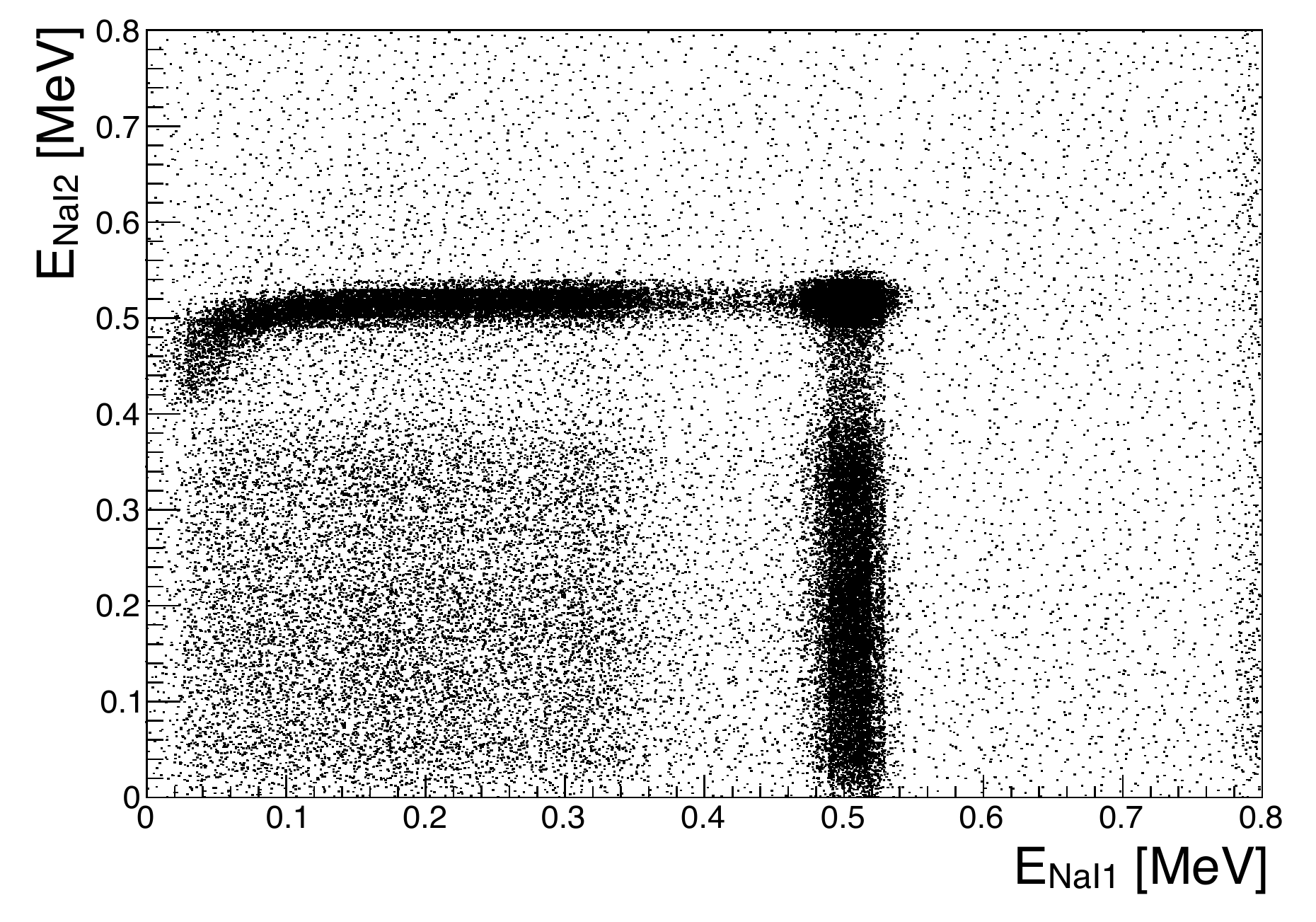}
\includegraphics [width =0.49 \textwidth] {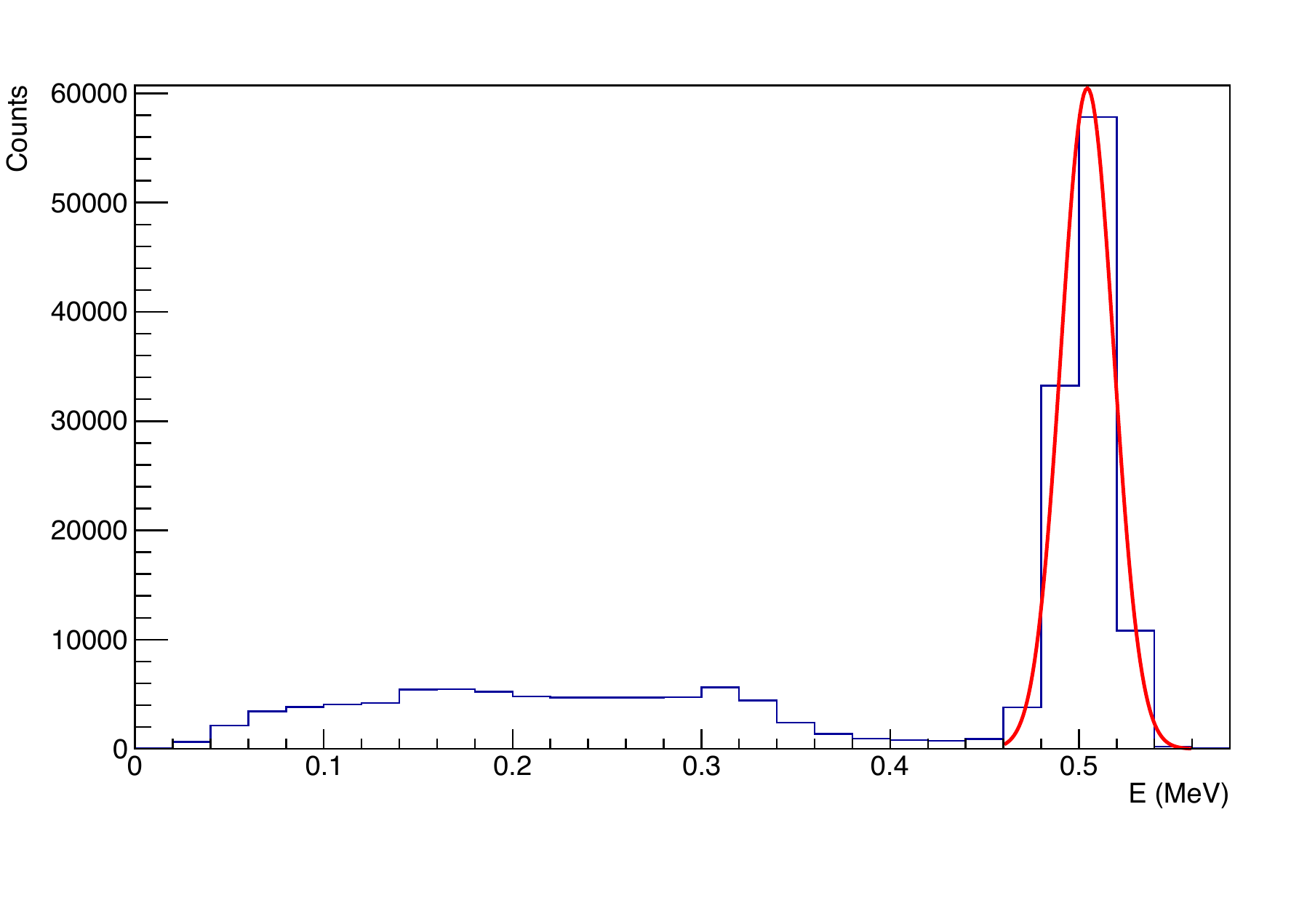}
\caption{\small{Left: correlation between the calibrated energy depositions in the two NaI detectors. Right: calibrated energy spectrum in both NaI detectors after requiring one of the two to be in the signal region ($mean=(0.507\pm0.012)\ \mega\electronvolt$).}}
\label{fig:NaI}
\end{center}
\end{figure}

Fig.~\ref{fig:NaI} shows the correlation between the $E_{\gamma}^1$ and $E_{\gamma}^2$ and the energy spectrum of both detectors after  requiring the other one to be in the signal window. 
A $\beta^+$ decay is identified by the presence of two back-to-back photons with an energy compatible with $E=511\ \kilo\electronvolt$. Thus the events with both energy depositions  in the signal window provide a background free measurement of the number of $\beta^+$ emitters. 
 
To evaluate the detection efficiency  ($\epsilon_{det}$) we simulated the experimental setup with GATE~\cite{Gate},  a simulation tool dedicated to medical imaging, radiontherapy and hadrontherapy based on  the GEANT4 MC code~\cite{Geant1,Geant2}. 

\begin{figure}[!ht]
\begin{center}
\includegraphics [width =0.5 \textwidth] {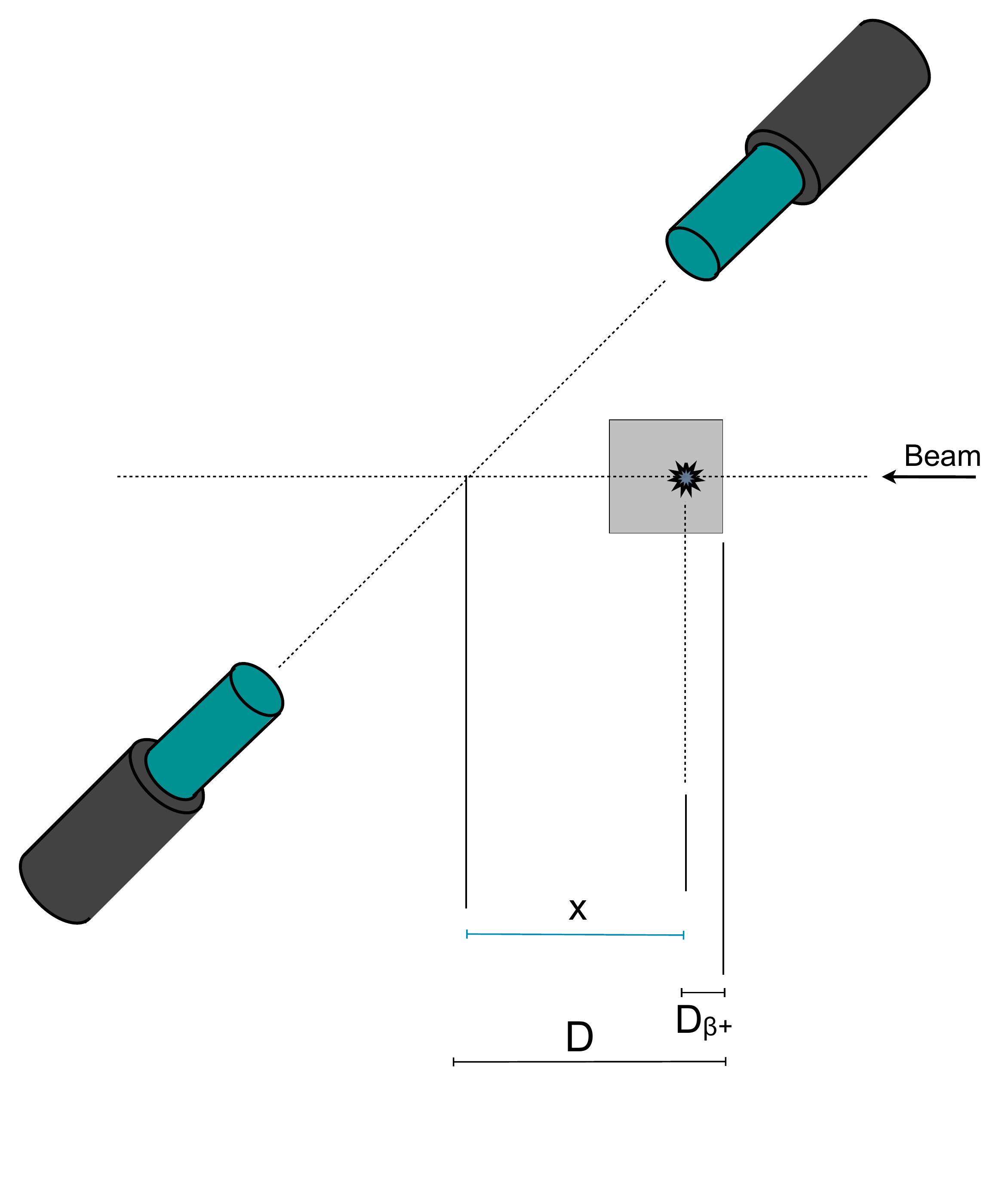}
\caption{\small{Definition of the $x$ and the $D$ variables (see text).}}
\label{fig:X}
\end{center}
\end{figure}

In order to measure the position of the peak of the $\beta^+$ decay spatial distribution it is important to evaluate the dependence of $\epsilon_{det}$ on the position of the two photon vertex with respect to the NaI detectors. To this aim we define $x$ as the distance along the beam line between the two photon vertex in the PMMA and the interception between the beam line and the line connecting the two detectors (see Fig.~\ref{fig:X}).  

\begin{figure}[!ht]
\begin{center}
\includegraphics [width =0.8 \textwidth] {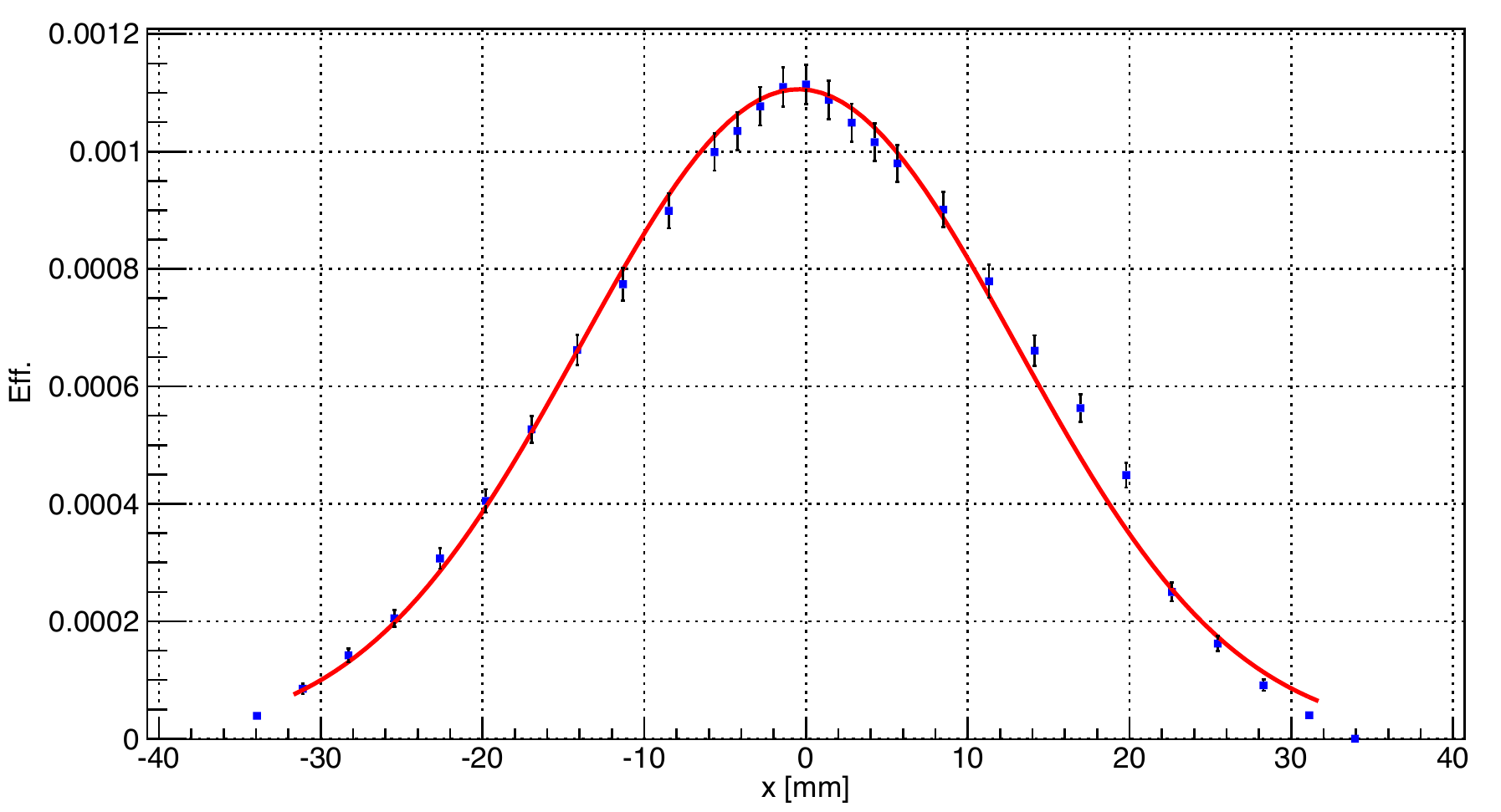}
\caption{\small{Detection efficiency as a function of the PMMA position on the beam line ($\chi^2/ndf=74/24$, $x_{mean}=(-0.046\pm0.13)\ \milli\meter$ and $\sigma=(13.45\pm0.09)\ \milli\meter$).  See text for details.}}
\label{fig:EffMC}
\end{center}
\end{figure}

The resulting functional dependence $\epsilon_{det}(x)$ is shown in Fig.~\ref{fig:EffMC}  and was parametrized with a single gaussian.\\ 

In order to measure the $\gamma-$PET rate, the number of carbon ions $N_C$ must be evaluated. In a given time interval, $N_C$ is computed by counting the number of signals given by the Start Counter ($N_{SC}$) within randomly-triggered time-windows of $T_w=2\ \micro\second$. From the number of time windows considered ($N_w$) and the total acquisition time ($T_{tot}$), the number of carbon ions $N_C$ is estimated to be: 

\begin{equation}
N_C=\frac{N_{SC}}{\epsilon_{SC}}\frac{T_{tot}}{N_wT_w}.
\end{equation}

The Start Counter efficiency $\epsilon_{SC}=(96\pm 1)\%$ has been estimated by exploiting the  two-sided PMT readout with negligible dark counts.

The number of impinging carbon ions has to be corrected for the dead-time inefficiency, $\epsilon_{DT}$ has been estimated from the total acquisition dead time ($T_{dead}$) as:
\begin{equation}
\epsilon_{DT}=1-\frac{T_{dead}}{T_{tot}}.
\end{equation}
The measured values of $\epsilon_{DT}$ range from  $70\%$ at an average carbon ion rate of $0.6\ \mega\hertz$ up to $47\%$ at $2\ \mega\hertz$. This efficiency correction was then applied to data as described below.

\section{Measurements of the time evolution of the $\beta^+$ emission}
\label{decadimento}

The rate of $\beta^+$ decays and the isotopic composition of the emitters was measured as a function of time both during irradiation and in the intervals in between. The time dependence of the emission during the irradiation results from two main contributions: ($i$) the creation of new emitters induced by the passage of the carbon ions in the PMMA, and ($ii$) the decay of the previously created ones. When the irradiation time is comparable to the decay time of the emitters, the relation between the emitter and dose rates is non-trivial. This is the case studied in this paper.

\subsection{Isotopic composition of emitters}
\label{sec:isotopic}

\begin{figure}[!ht]
\begin{center}
\includegraphics [width =0.65 \textwidth] {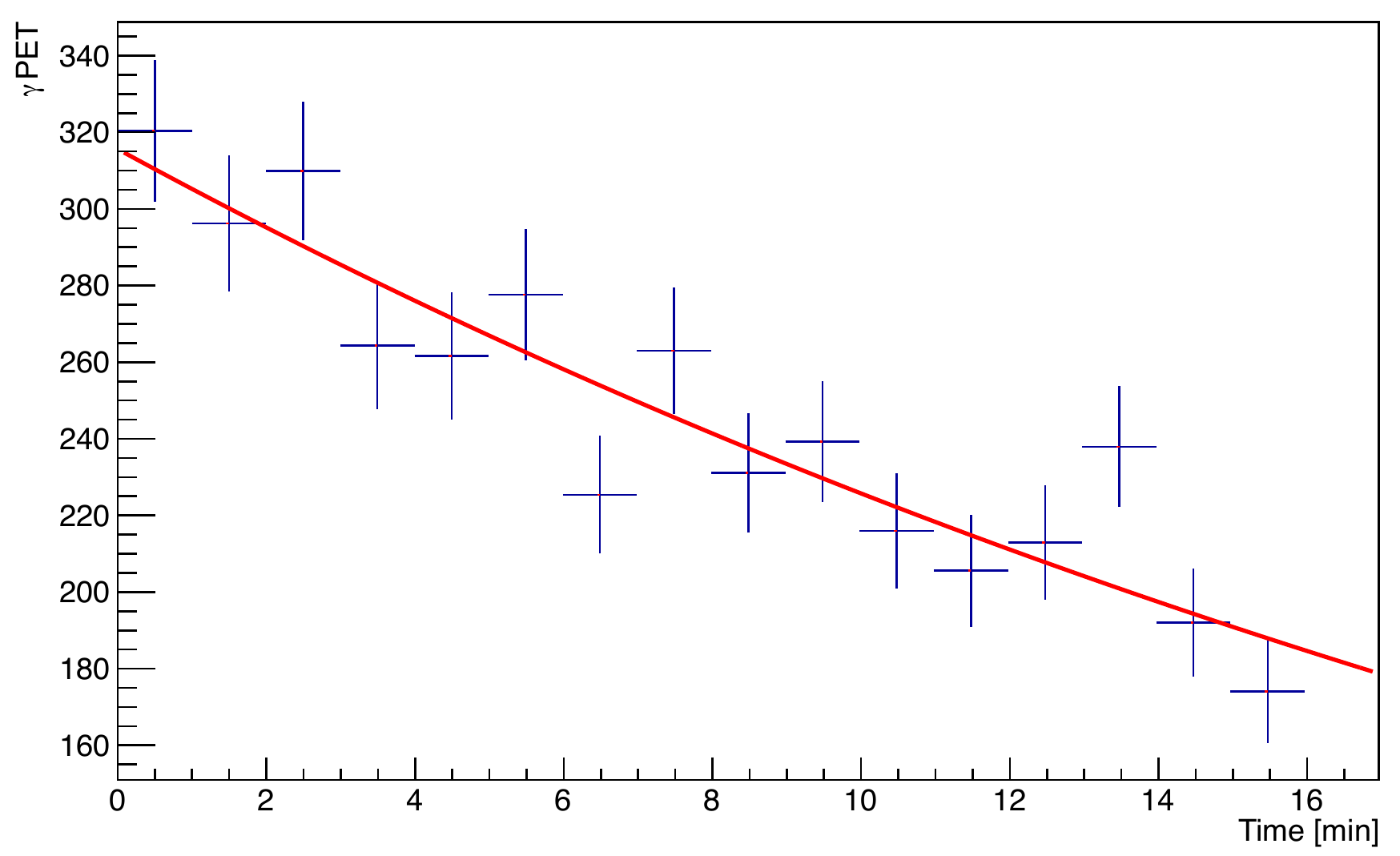}
\caption{\small{Distribution of the number of detected $\gamma-$PET as a function of time in a run with no-beam; the fit with an exponential function is shown ($\tau = 33 \pm 3\ \minute$, $\chi ^2 /ndof = 15.3/13$).}}
\label{fig:PETboff}
\end{center}
\end{figure}

The isotopes that can be produced during the carbon ion irradiation of the PMMA are $^{11}C$, $^{13}N$,  $^{15}O$, $^{14}O$ and $^{10}C$ with half-lives of $29\ \minute$, $15\ \minute$,  $2\ \minute$, $100\ \second$ and $19\ \second$ respectively. The wide spread of lifetimes allows to discriminate among isotopes. We chose those isotopes because they are the one with longer half-lives and, as it will be seen later on in the ``during irradiation'' measurements, the gamma distribution is time has a rise of the order of tens of minutes. When the beam is turned off (t=0), the time evolution of the $\beta^+$ decay rate ($R(t)$), is: 

\begin{equation}
\label{eq:decay}
R(t)=R(0)\sum_{i=1}^{N_s} f_i e^{-t/\tau_i}.
\end{equation}

assuming $N_s$ different species of isotopes with fractions $f_i$ and lifetimes $\tau_i$. The measured $\gamma-$PET, expected to be proportional to $R(t)$, in a run with no-beam (measurements after irradiation) as a function of time is shown in Fig.~\ref{fig:PETboff}. 

Since the time $t$ is measured with respect to the start of the run and not to the end of the irradiation as required by Eq.~\ref{eq:decay},  the sensitivity to the fast decaying isotopes ($^{15}O$ and  $^{14}O$) and the combination of different runs are compromised. We have therefore decided to analyze only the no-beam run with the largest statistics (shown in Fig.~\ref{fig:PETboff}). The available data does not allow us to perform a fit with a double exponential function accounting for both $^{11}C$ and $^{13}N$ populations and therefore we can only test the hypothesis that one isotope dominates. The fit with a single exponential function results in a lifetime $\tau = 33 \pm 3\ \minute$ ($\chi ^2 /ndof = 15.3/13$), thus indicating that the $^{11}C$ component  is dominant.

\subsection{Time evolution of the emission rate during irradiation}
\label{tdep}

As stated in the introduction of section~\ref{decadimento}, the time evolution of the  $\beta^+$ emitters becomes non trivial when irradiation lasts for a time interval comparable with the lifetime of the emitters. We therefore elaborated a model to describe the simultaneous occurrence of new activations and decays of the $\beta^+$ emitters and used it to fit to the time evolution of the measured rates, testing its goodness and extracting its parameters.\\
Defining the number of $\beta^+$ emitters for each of $N_s$ species as $N_{dec}^s$ ($s=1,...,N_s$), with $\tau_s$ lifetime, we can write:

\begin{equation}
\frac{dN_{dec}^s(t)}{dt}=A_s \frac{dN_C (t)}{dt} - \frac{1}{\tau_s}N_{dec}^s(t).
\label{eq:Nevol}
\end{equation}

The first term on the right side of the Eq.~\ref{eq:Nevol} represents the activation induced by the impinging carbon ions, with  $A_s^{raw}$ representing the fraction of carbon ions activating $\beta^+$ emitters. The second term takes into account the  decay of the activated nuclei with decay times $\tau_s$.  Each species has an independent time evolution, so the measured $\gamma-$PET evolution in the detector ($N_{\gamma}$) takes into account all species contributions:

\begin{equation}
\frac{dN_{\gamma}(t)}{dt}=\epsilon_{det}\sum_{s=1}^N\frac{N_{dec}^s(t)}{\tau_s}.
\label{eq:gamma}
\end{equation}

\begin{figure}[!ht]
\begin{center}
\includegraphics [width =0.48 \textwidth] {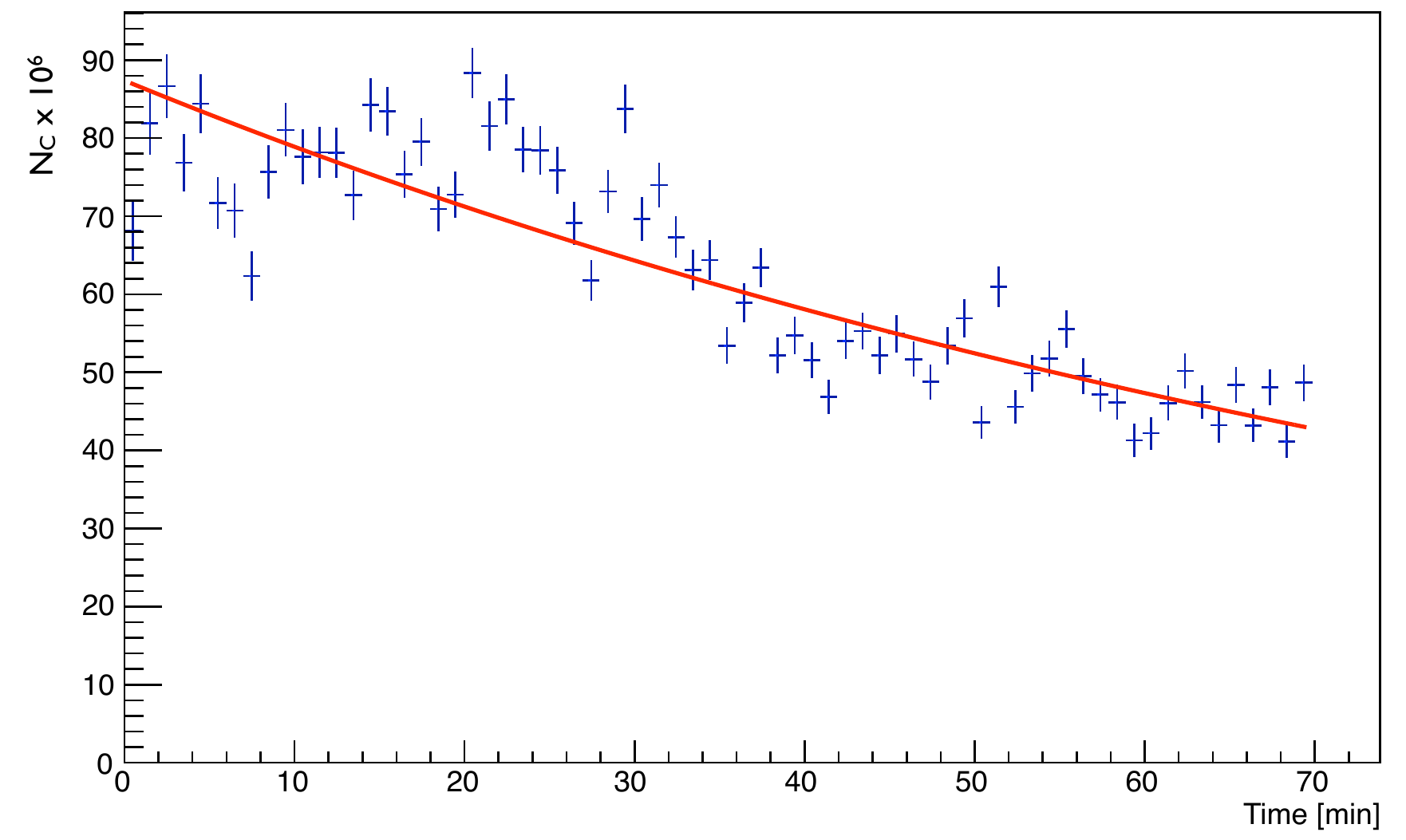}
\includegraphics [width =0.48 \textwidth] {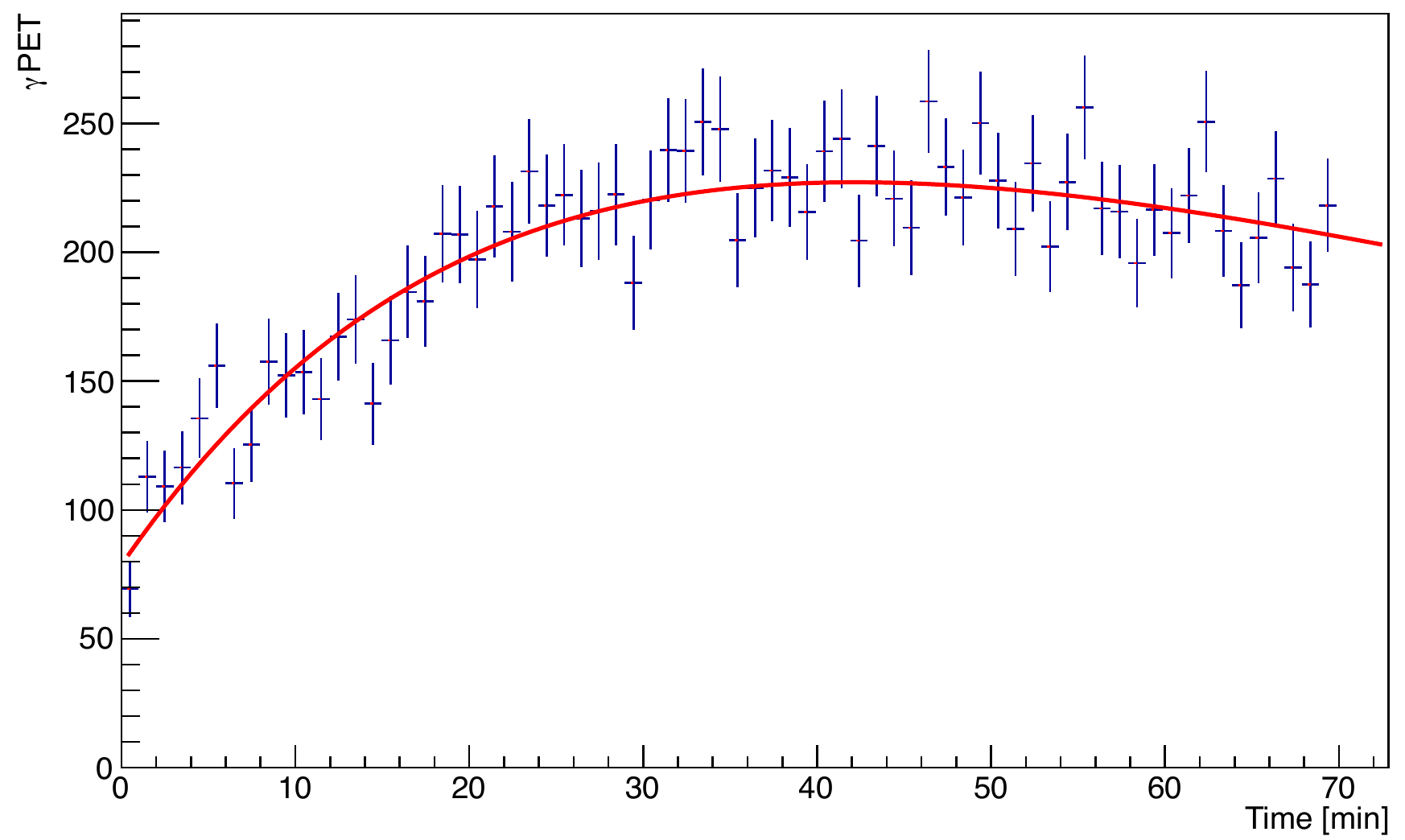}
\caption{\small{Measured $N_c$ (Left) and  $N_{\gamma}$ (Right) as a function of time. The fit of the measured $N_{\gamma}$ distribution using Eq.~\protect\ref{eq:esplicita} (solid line) gives the species ratio $^{11}C$ over $^{13}N$ (see Eq.~\protect\ref{eq:AcAn}).}}
\label{fig:PETbon}
\end{center}
\end{figure}

Eq.~\ref{eq:Nevol} requires the knowledge of $N_C(t)$. As it can be seen for two different runs shown in Figs.~\ref{fig:PETbon} (Left) and ~\ref{fig:PETnotte} (Left), $N_C(t)$ behavior cannot be always described with a function that allows Eqs.~\ref{eq:Nevol} to be solved. We therefore developed two different methods: ($i$) an analytical one to be used when $N_C(t)$ can be described by an analytical function, and ($ii$) a numerical one that can always be applied, but requires the assumption of one dominant isotope among the emitters. It is to be noted that accounting for the time dependence of the beam intensity is particularly needed when the irradiation lasts for periods comparable to the decay times of the isotopes, as it is the case in the therapeutical treatments.

\subsubsection{Analytical method}

The Fig.~\ref{fig:PETbon} (Left) shows the measured $N_C(t)$ fit by an exponential function ($N_C(t)=\nu_{beam} \ exp(- t / \tau_{beam})$). In this case, the solution of Eqs.~\ref{eq:Nevol} and ~\ref{eq:gamma} is: 

\begin{equation}
N_{\gamma}=\epsilon_{det}\sum_{s=1}^{N_s} \left(\left(\frac {A_s\tau_s}{\frac{1}{\tau_s}-\frac{1}{\tau_{beam}}}\right)  \nu_{beam} \ e^{-\frac{t}{\tau_{beam}}} + \frac{\alpha_s}{\tau_s}\ e^{-\frac{t}{\tau_s}} \right).
\label{eq:Ngamma}
\end{equation}

with 

$$
\alpha_s=-\epsilon_{det}\nu_{beam}\frac{A_s}{\frac{1}{\tau_{s}}-\frac{1}{\tau_{beam}}}e^{\left(\frac{1}{\tau_s}-\frac{1}{\tau_{beam}}\right) t_0}.
$$

In Sec.~\ref{sec:isotopic} the lifetime measurement obtained from the fit of Fig.~\ref{fig:PETboff} indicates that the $^{11}C$ component dominates. By using Eq.~\ref{eq:Ngamma} we can now measure the relative contribution to $N_{\gamma}$ from $^{11}C$ and $^{13}N$ species, $A_C$ and $A_N$. 
Assuming $t_0$ as the time for which $N_{dec}^C(t_0)=N_{dec}^N(t_0)=0$, Eq.~\ref{eq:Ngamma} can be written as follow: 

\begin{equation}
N_{\gamma}=\epsilon_{det}A_N\left(\frac{1 / \tau_N}{\frac{1}{\tau_N}-\frac{1}{\tau_{beam}}}+\frac{A_C / (A_N \cdot \tau_C)}{\frac{1}{\tau_C}-\frac{1}{\tau_{beam}}}\right)\nu_{beam} \left(e^{-\frac{t}{\tau_{beam}}}-e^{-\frac{t_0}{\tau_{beam}}} \right ).
\label{eq:esplicita}
\end{equation}

Fitting the time distribution of $N_{\gamma}$ with Fig.~\ref{fig:PETbon} (Right), we obtain:

\begin{equation}
\frac{A_C}{A_N} = 16.6 \pm 2.7.
\label{eq:AcAn}
\end{equation} 

The dominance of  $^{11}C$ over $^{13}N$ is therefore confirmed and the measured ratio is consistent with the results, obtained with a higher energy beam, in Ref.~\cite{rates2009}.

\subsubsection{Numerical method}

\begin{figure}[!ht]
\begin{center}
\includegraphics [width =0.8 \textwidth] {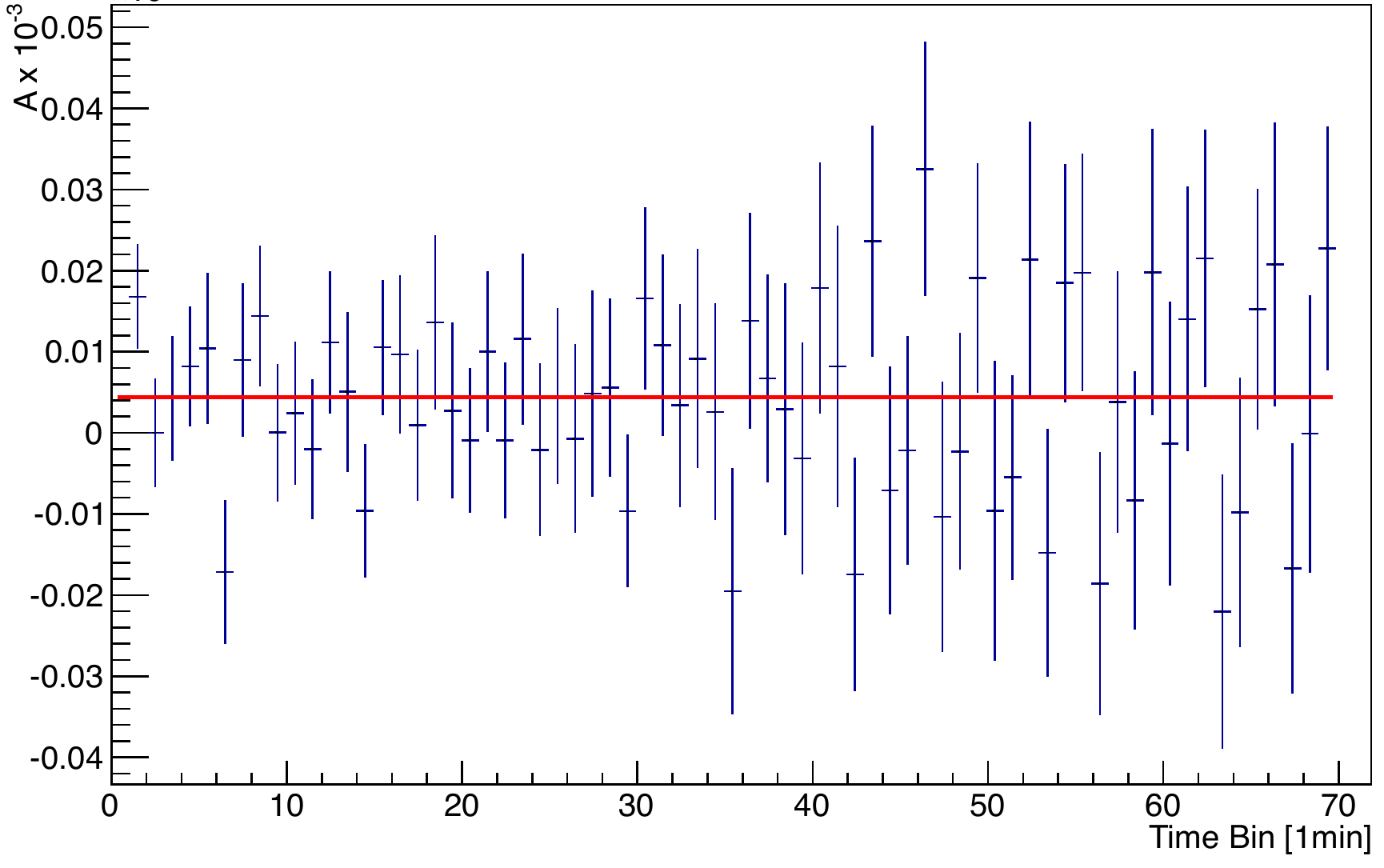}
\caption{\small{$A^{raw}$ parameter time evolution and fit to a constant.}}
\label{fig:ParA}
\end{center}
\end{figure}

In general the carbon ion rate cannot be parametrized with a simple function allowing an analytical solution of Eq.~\ref{eq:Nevol}. Therefore, a numerical method valid in the hypothesis of one dominant isotope only has been developed.  
Defining the integral of the number of carbon ions ($N_C^i$) and of the detected $\gamma-$PET ($N_{\gamma}^i$) in the time bin $i$  with $B_W$ width, we can estimate, by  approximating the derivative with the differential increment,  the $A^{raw}$ parameter in each single bin:

\begin{equation}
A_i^{raw}= \frac{1 }{N_C^i}\cdot \left( \frac{N_{\gamma}^i}{\epsilon^i_{DT}} + \frac{\tau}{B_W}\cdot \left(  \frac{N_{\gamma}^i}{\epsilon^i_{DT}} - \frac{N^{i-1}_{\gamma}}{\epsilon^{i-1}_{DT}} \right) \right).
\label{eq:A}
\end{equation}

where the $raw$ suffix indicates that the measurement has not been corrected for the detector efficiency. As an example, the numerical method has been applied to the $N_C$ and $N_{\gamma}$ measurements shown in Fig.~\ref{fig:PETbon}. 
The measured $A^{raw}_i$ parameters using Eq.~\ref{eq:A} are reported in Fig.~\ref{fig:ParA}.
By minimizing the $\chi^2$, defined taking into account correlations and assuming that the measured parameter is independent of the time bin, we obtain a mean value $A_m^{raw}=(4.7\pm1.0) \cdot 10^{-6}$, with $\chi^2/DOF=37/69$.
 
Conversely, we can exploit the same equations (\ref{eq:A}), assuming a measurement of  $A^{raw}_m$  from a calibration run, to  estimate the number of incident carbon ions ($N_C^{\gamma}$), once the numbers $N^{i}_{\gamma}$ (Eq.~\ref{eq:A}) are measured for a specific data sample:

\begin{equation}
N_C^{\gamma}= \frac{1 }{A^{raw}_m}\sum_{i=1}^{N_{bin}} \left( \frac{N_{\gamma}^i}{\epsilon^i_{DT}} + \frac{\tau}{B_W}\cdot \left(  \frac{N_{\gamma}^i}{\epsilon^i_{DT}} - \frac{N^{i-1}_{\gamma}}{\epsilon^{i-1}_{DT}} \right) \right).
\label{eq:Nc}
\end{equation}

\begin{figure}[!ht]
\begin{center}
\includegraphics [width =0.46 \textwidth] {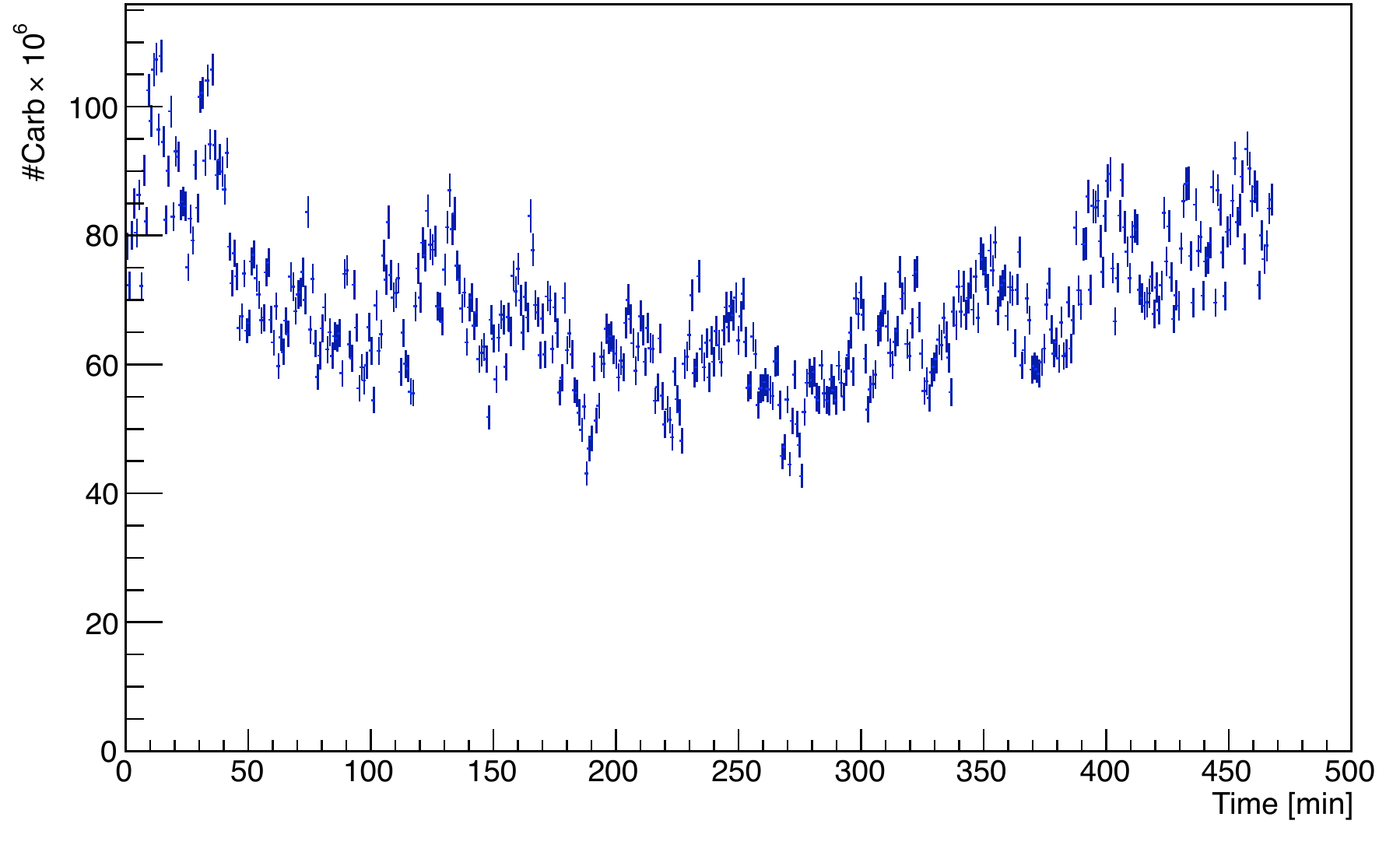}
\includegraphics [width =0.49 \textwidth] {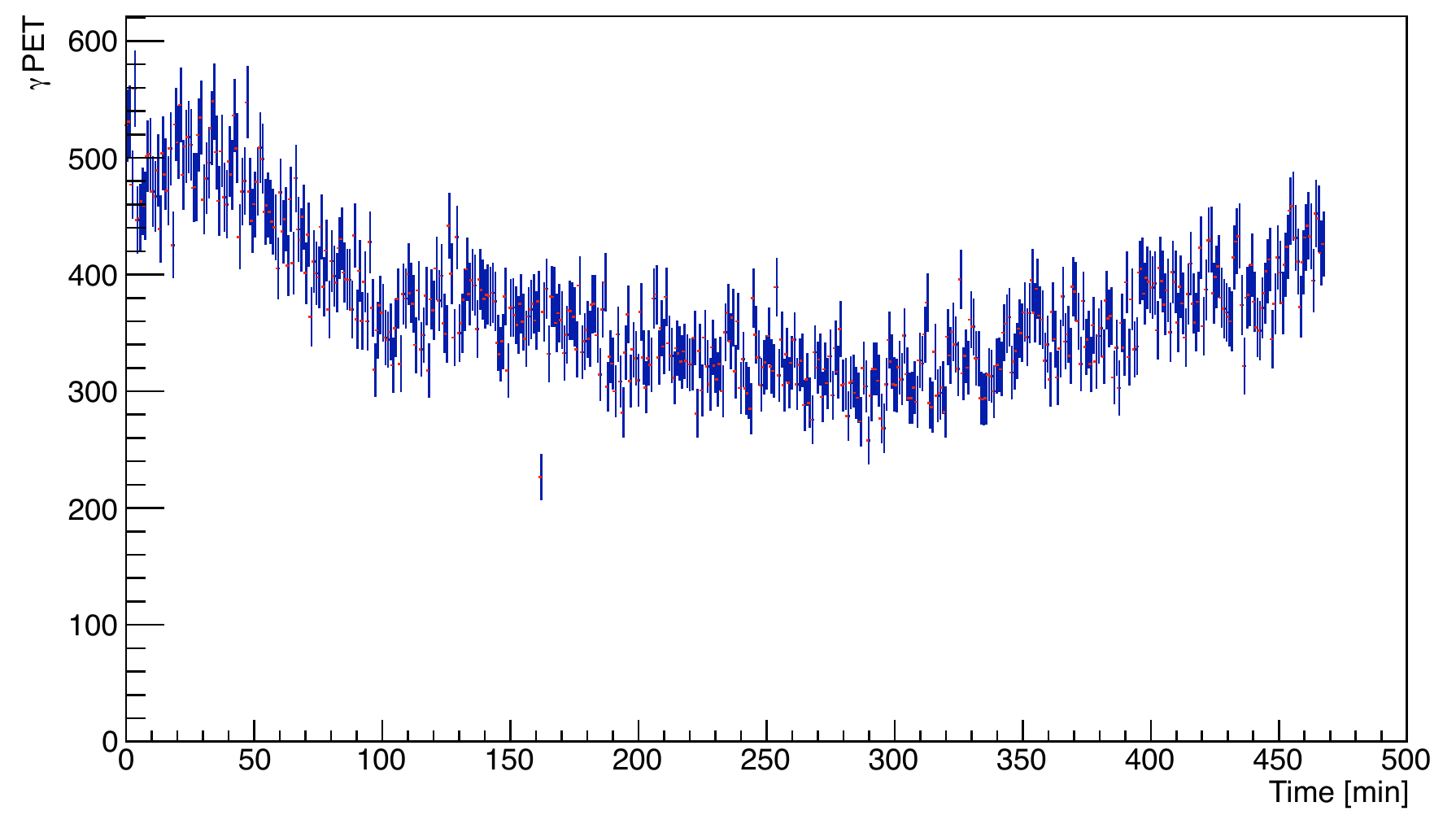}
\caption{\small{Measured $N_c$ (Left) and $N_{\gamma}$ (Right) as a function of time.}}
\label{fig:PETnotte}
\end{center}
\end{figure}

We have tested this procedure on a long run (see Fig.~\ref{fig:PETnotte} for the measured  $N_{\gamma}$ and $N_C$) with slightly unstable beam conditions. The $A_m^{raw}$ parameter was estimated on the shorter run from which Fig.~\ref{fig:ParA} was taken, after adjusting for the detector efficiency as detailed in the next section.

\begin{figure}[!ht]
\begin{center}
\includegraphics [width =1 \textwidth] {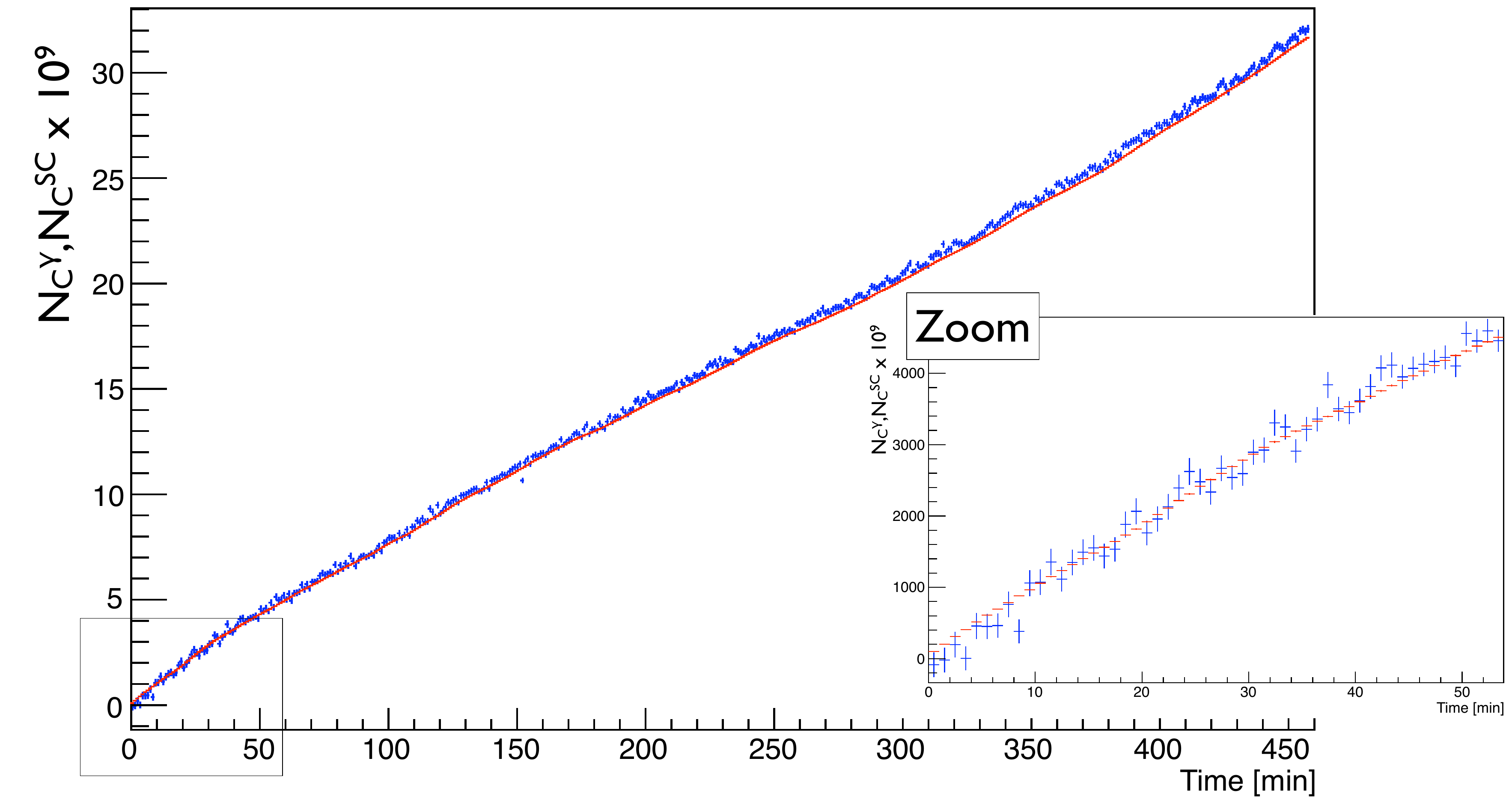}
\caption{\small{Cumulative distribution of the number of carbon ion measured with the Start Counter, $N_C^{SC}$ (magenta dashed data), compared to the number of ions estimated by the measurements of $\gamma-PET$, $N_C^{\gamma}$ (blue dot points). The plot on the right is a zoom in the first $50\ \minute$ of acquisition. }}
\label{fig:Nc_sc}
\end{center}
\end{figure}

Fig.~\ref{fig:Nc_sc} shows the cumulative distribution of the number of ions estimated by the measurements of $\gamma-PET$, $N_C^{\gamma}$, blue dot points, compared to the cumulative number of carbon ions measured with the Start Counter $N_C^{SC}$, magenta dashed data.  A good agreement is visible,  also at times comparable with the lifetimes of the decaying isotopes. 

\section{Spatial dependence of the $\beta^+$ emission}
\label{Bragg}

The proposed numerical method to estimate $N_C$ relies on the knowledge of the $A_m^{raw}$ parameter (see Eq.~\ref{eq:Nc}). The $A_m^{raw}$ parameter measurement described so far was corrected only for the dead-time efficiency. To obtain the final $A$ parameter we must correct for the detector efficiency ($\epsilon_{det}$). The latter depends on the position of the $\beta^+$ emitters along the beam line with respect to the line connecting the two NaI detectors, as discussed in Sec.~\ref{Eff}. In order to study the spatial distribution of the emitters the  $A_m^{raw}$ parameter was measured for several positions of the PMMA, i.e. for several values of $D$ as defined in Fig.~\ref{fig:X}. It can be written as:

\begin{equation}
A_m^{raw}(D)=\int \ \epsilon_{det} (x) \ \frac{dA_{ec}}{dx}(x|D){dx}
\label{eq:Acorr}
\end{equation}
%questo $A_ec$ non si capisce molto bene cosa sia
with $dA_{ec}/ dx(x|D)$ the density of emitters as a function of the depth $x$  in the PMMA (as defined in Fig.~\ref{fig:X}). The dependence on $D$ takes into account the distribution shown in Fig.~\ref{fig:Beta} and the relative position of the PMMA and the NaI detectors. Since the literature~\cite{Bragg} reports widths of the $\beta+$ emitter peak distribution smaller than $\epsilon(x)$ (comparison of Fig.~\ref{fig:Beta} and Fig.~\ref{fig:EffMC}), we approximate $dA_{ec}(x|D)/ dx\sim\delta(x-(D-D_{\beta^+}))$ where $D_{\beta^+}$ is the position of the $\beta^+$ emitters within the PMMA. Finally we  obtain:

\begin{equation}
A_m^{raw}(D)=A_0\cdot \epsilon_{det}(D-D_{\beta^+})
\label{eq:fitfun}
\end{equation}
with $A_0$ the constant component of the $A$ parameter, $D$ the distance along the beam line between the line connecting the NaI crystals and the beam entrance face of the PMMA  (see Fig.~\ref{fig:X}) and $D_{\beta^+}$ the mean position of the emitters in the same reference system.

\begin{figure}[!ht]
\begin{center}
\includegraphics [width =0.8 \textwidth] {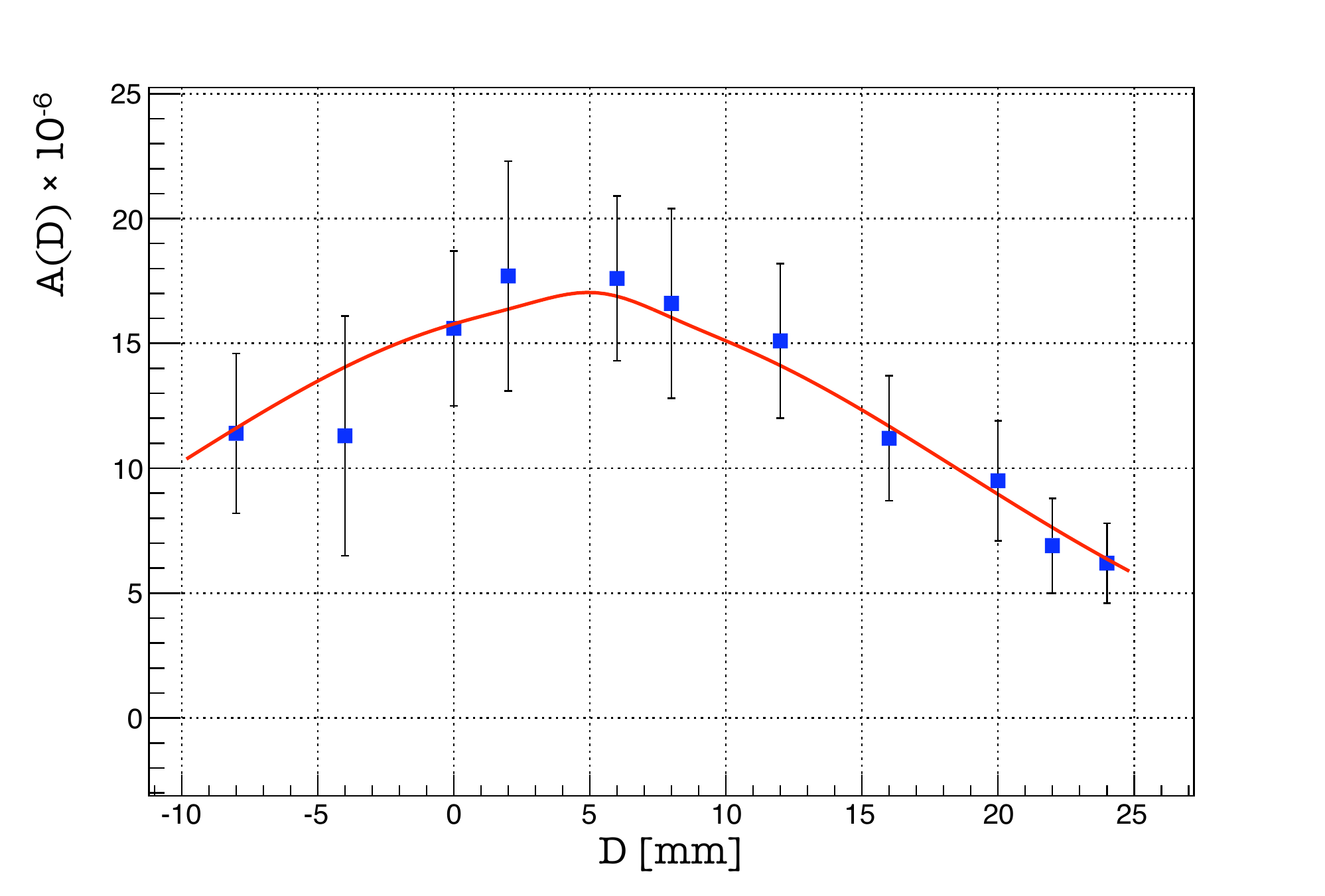}
\caption{\small{The measured $A_m^{raw}(D)$ parameter as a function of the position of the PMMA, fitted as described in the text.}}
\label{fig:Afit}
\end{center}
\end{figure}

Fig.~\ref{fig:Afit}  shows the dependence of the measured $A$ parameter from $D$: the width of the distribution %($\sim 12.5\ \milli\meter$)
  is consistent with the one of the efficiency in Fig.~\ref{fig:EffMC} 
 %($13.5\ \milli\meter$) 
 thus confirming the hypothesis used to obtain Eq.~\ref{eq:fitfun}.

From the fit of the spectrum in Fig.~\ref{fig:Afit} with the function in Eq.~\ref{eq:fitfun} we obtain:

 \begin{eqnarray}
 A_0 &=& (10.3 \pm 0.7) \cdot 10^{-3}\\
 D_{\beta+}&=&(5.3 \pm 1.1)\ \milli\meter.
 \label{eq:beta+}
\end{eqnarray}
 
\begin{figure}[!ht]
\begin{center}
\includegraphics [width =0.8 \textwidth] {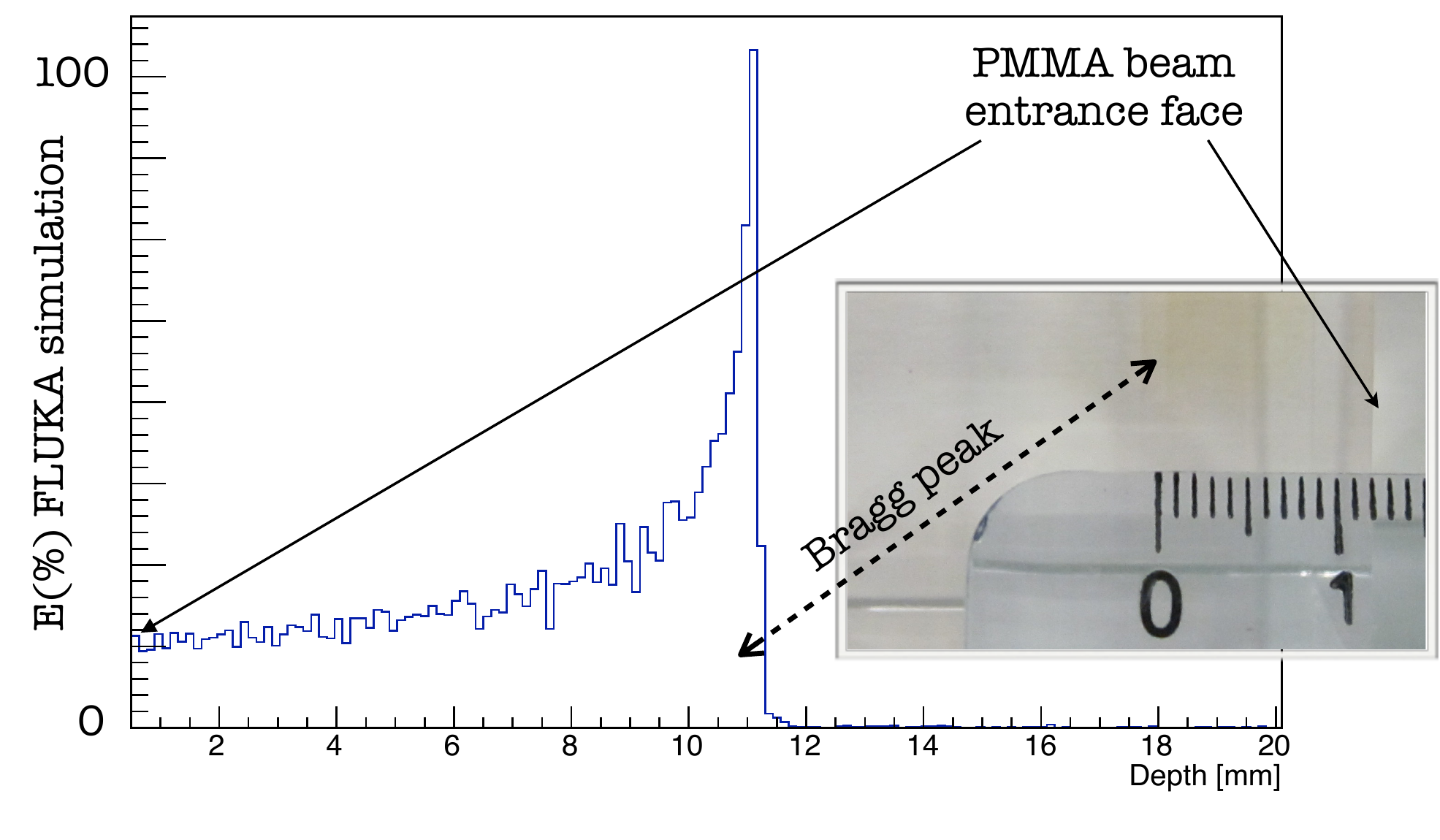}
\caption{\small{Dose deposition in the PMMA simulated with FLUKA as a function of the depth from the beam entrance face of the PMMA.}}
\label{fig:Fluka}
\end{center}
\end{figure}

The dose deposition in the PMMA has been simulated with FLUKA~\cite{FLUKA}. The result is shown in Fig.~\ref{fig:Fluka} as a function of the depth from the beam entrance face of the PMMA with the beam entering form the left side of the plot. In our configuration the Bragg peak is at $11.0 \pm 0.5\ \milli\meter$  from the beam entrance face (light yellow band) of the target. The simulation is confirmed by the picture of the PMMA after the data taking , shown in Fig.~\ref{fig:Fluka} (Inset): the dose distribution is visible by the deterioration (light yellow band) of the PMMA.

Using the measured value of $D_{\beta+}$ (Eq.~\ref{eq:beta+}) with the simulated Bragg peak position $D_{Bragg}$, we obtain the distance $\xi$ between the $\beta+$ emitters and the Bragg peaks: $\xi = | D_{\beta+} - D_{Bragg} |= (5.7 \pm 1.2)\ \milli\meter$.

\section*{Conclusions} 

We presented a study of the rate of $\gamma -$PET produced in the  interaction of $80\ \mega\electronvolt/$u fully stripped carbon ions with a PMMA phantom, using a pair of NaI crystal detectors.

We proposed, and validated with data, a model to describe the activated nuclei $\beta^+$ decay during the irradiation. We demonstrated the possibility to estimate the number of impinging carbon ions from the number of observed $\gamma-$PET.

We measured the ratio between the number of activated  $^{11}C$ and $^{13}N$ to be $A_C / A_N = 16.6 \pm 2.7$ and a number of $(10.3 \pm 0.7) \cdot 10^{-3}$  generated $^{11}C$ ions, per impinging carbon ion undergoing $\beta^+$ decay.

Finally we measured the mean position of the $\beta^+$ emission  to be  $ D_{\beta+}=(5.3 \pm 1.1)\ \milli\meter$ from the beam entrance to the PMMA face, to be compared to the simulated Bragg peak position  $D_{Bragg}=(11.0\pm 0.5)\ \milli\meter$. Of course more accuracy on the measurement could be achieved with larger scale experiments, but anyhow we think that fair information can also be achieved from the comparison with the presented data.

All this information can be used as a benchmark for the $\beta^+$ emitters MC simulation of hadrontherapy.

\begin{center}\textbf{\large Acknowledgements}\end{center}{\large \par}

We would like to thank Carmelo Piscitelli for the realization of the mechanical support. The staff of the INFN-LNS (Catania, Italy) test beam is gratefully acknowledged for their kind cooperation and helpfulness.


\begin{thebibliography}{1}
\bibitem{Bragg}
I.~Pshenichnov {\em et~al.}, {\it Distributions of positron-emitting nuclei in proton and carbon ion therapy studied with GEANT4} (2006)
Physics in Medicine and Biology, Volume $51$, Number $23$.

\bibitem{Amaldi}
U.~Amaldi and G.~Kraft, {\it {Radiotherapy with beams of carbon ions}},  \\
{\em Reports on Progress in Physics} {\bf 68} (2005), no.~8 1861.


\bibitem{inBeam:th}
J. Pawelke {\em et.~al.} {\it{In-beam PET imaging for the control of heavy-ion tumour therapy.}},  IEEE Trans. Nucl. Sci.{\bf 44} (1997) 1492.

\bibitem{inBeam:C}
F. Fiedler {\em et.~al.} {\it{In-beam PET measurements of biological half-lives of $^{12}$C irradiation induced $\beta^+$-activity.}}, Acta onc. Stockholm Sweden  {\bf 47} (2008) 1077.

\bibitem{inBeam:exp}
K. Parodi {\em et.~al.} {\it{In-beam PET measurements of $\beta^+$ radioactivity induced by proton beams.}}, Phys. Med. Biol. {\bf 47} (2002) 21; 
F. Fiedler  {\em et.~al.}, {\it The feasibility of in-beam PET for therapeutic beams of He-3},  IEEE Trans. Nucl. Sci. {\bf 53} ( 2006) 2252;
  S. Vecchio {\em et.~al.} {\it{A PET prototype for in-Beam monitoring of proton therapy}},  IEEE Trans. Nucl. Sci.{\bf 56} (2009) 1;
M. Priegnitz {\em et.~al.} {\it{An Experiment-Based Approach for Predicting Positron Emitter Distributions Produced During Therapeutic Ion Irradiation }},  IEEE Trans. Nucl. Sci.{\bf 59} (2012) 77;
\bibitem{rates2009} N. Sommerer  {\em et.~al.} {\it{In-beam PET monitoring of mono-energetic (16)O and (12)C beams: experiments and FLUKA simulations for homogeneous targets }}, Phys. Med. Biol. {\bf 54} (2009) 3979 ; 


\bibitem{PaperoLYSO}
C.~Agodi  {\em et.~al.}, {\it {Precise measurement of prompt photon emission for carbon ion
  therapy}}, JINST {\bf 7} (2011) 03001
  
\bibitem{Gate}
S.~Jan {\em et.~al.}, {\it Gate: a simulation toolkit for pet and spect},  {\em
  Phys. Med. Biol.} {\bf 49} (2004) 4543.

\bibitem{Geant1}
S.~Agostinelli {\em et.~al.}, {\it G4--a simulation toolkit},  {\em Nuclear
  Instruments and Methods in Physics Research Section A: Accelerators,
  Spectrometers, Detectors and Associated Equipment} {\bf 506} (2003), no.~3
  250.

\bibitem{Geant2}
J.~Allison {\em et.~al.}, {\it Geant4 developments and applications},  
  IEEE Trans. Nucl. Sci. {\bf 53} (2006) 270.

\bibitem{FLUKA}
A.~Fasso, A.~Ferrari, S.~Roesler, P.~Sala, F.~Ballarini, {\em et.~al.}, {\it
  {The Physics models of FLUKA: Status and recent developments}}, {\em arXiv:hep-ph/0306267} (2003)


\end{thebibliography}
\end{document}